\documentclass[12pt,reqno]{amsart}
\usepackage{amsmath,amsfonts,amsthm,amssymb,dsfont}
\newcommand\version{August 13, 2020}
\usepackage{amsxtra}

\newtheorem{theorem}{Theorem}[section]

\newtheorem{lemma}{Lemma}[section]

\newtheorem{conjecture}{Conjecture}[section]

\theoremstyle{definition}

\theoremstyle{remark}

\numberwithin{equation}{section}

\newcommand{\C}{\mathbb{C}}

\newcommand{\eps}{\epsilon}
\renewcommand{\epsilon}{\varepsilon}

\newcommand{\F}{\mathcal{F}}

\newcommand{\N}{\mathbb{N}}

\renewcommand{\phi}{\varphi}
\newcommand{\R}{\mathbb{R}}
\newcommand{\Sph}{\mathbb{S}}

\newcommand{\Hh}{\mathfrak{H}}
\newcommand{\ch}{\mathord{\mathfrak h}}
\newcommand{\E}{\mathcal{E}}
\newcommand{\M}{\mathcal{M}}
\newcommand\infspec{{\rm inf\, spec\, }}
\newcommand\id{{\mathds{1}}}
\newcommand\const{{\rm const. \, }}

\DeclareMathOperator{\Tr}{Tr}

\newcommand{\beq}{\begin{equation}}
\newcommand{\eeq}{\end{equation}}

\begin{document}

\title{The Polaron at Strong Coupling}

\author{Robert Seiringer}
\address{IST Austria, Am Campus 1, 3400 Klosterneuburg, Austria}
\email{rseiring@ist.ac.at}

\begin{abstract} 
We review old and new results on the Fr\"ohlich polaron model. The discussion includes the validity of the (classical) Pekar approximation in the strong coupling limit, quantum corrections to this limit, as well as the divergence of the effective polaron mass.
\end{abstract}

\date{\version}

\maketitle

\section{Introduction}

\subsection{The Fr\"ohlich Polaron Model}

The physical system under consideration consists of a charged particle (e.g., an electron) interacting 
with the (quantized) phonons of a polar crystal \cite{AD}.  As the electron moves trough the medium, it induces 
a polarization field proportional to the electric field it creates. This polarization field, in turn, exerts a force on the electron.  

We consider in the following the case of a large polaron, where the de Broglie wave length of the electron is much larger than the lattice spacing in the medium, and hence the latter can be approximated by a continuum. The relevant mathematical model in this case is the  Fr\"ohlich Hamiltonian \cite{F37}. In order to describe it, recall the definition 
of the bosonic Fock space
\beq
\F = \bigoplus_{n=0}^\infty \Hh_n \quad , \quad \Hh_n = {\otimes_{\rm sym}^n} L^2(\R^3) 
\eeq
where ${\otimes_{\rm sym}^n} L^2(\R^3) $ consists of permutation-symmetric functions in $\otimes^n L^2(\R^3)$. All relevant operators on $\F$ can be expressed in terms of the 
creation and annihilation operators, defined as follows. For $\varphi \in \Hh_1= L^2(\R^3)$, $a(\varphi) : \Hh_n \to \Hh_{n-1}$ with
\beq
\left[ a(\varphi) \Psi_n \right] (x_1,\dots,x_{n-1}) = \sqrt{n} \int_{\R^3} \overline{\varphi(x_n)} \Psi(x_1,\dots,x_n) dx_n \,.
\eeq
Its adjoint $a^\dagger(\varphi) : \Hh_{n-1} \to \Hh_n$ is given by 
\beq
\left[ a^\dagger(\varphi) \Psi_{n-1}\right] (x_1,\dots,x_n) = \frac 1 {\sqrt{n}} \sum_{j=1}^n \varphi(x_j) \Psi(x_1,\dots, \, \not  \! x_j, \dots,x_n) \,.
\eeq
They satisfy the canonical commutation relations (CCR) 
\beq
[ a(\varphi), a^\dagger(\psi) ] = \langle \varphi | \psi \rangle_{\Hh_1} \quad , \quad [ a(\varphi), a(\psi) ] = 0 \,.
\eeq
Formally, one often writes
\beq
a^\dagger(\varphi) = \int_{\R^3} dx\, \varphi(x) a^\dagger_x  = \int_{\R^3}d k\, \hat \varphi(k) a^\dagger_k 
\eeq
for operator-valued distributions $a^\dagger_k$ satisfying
\beq
[ a_k , a^\dagger_{k'}] = \delta(k-k') \quad , \quad [ a_k , a_l ] =0 \,.
\eeq
Here $a^\dagger_k$ is really the Fourier transform of $a^\dagger_x$, but we shall sometimes use this sloppy notation when no confusion can arise. 

The number operator on $\F$ is simply given by $\N \Psi_n = n \Psi_n$ for $\Psi_n \in \Hh_n$. For an orthonormal bases $\{\varphi_j\}$ of $\Hh_1$, one can equivalently write 
\beq
\N = \sum_j a^\dagger(\varphi_j) a(\varphi_j) = \int_{\R^3} dk \, a^\dagger_k a_k \,.
\eeq

We consider here the Fr\"ohlich model \cite{F37} for one charged particle. The relevant Hilbert space in this case is $L^2(\R^3) \otimes \F$, and the Hamiltonian reads
\begin{equation}\label{def:ham}
H_\alpha = -\Delta - \sqrt{\frac{\alpha}{2\pi^2}}  \int_{\R^3}  \frac {dk} {|k|} \left( a_k e^{i k\cdot x} + a_k^\dagger e^{-ik \cdot x} \right) + \int_{\R^3} dk\, a_k^\dagger a_k 
\end{equation}
with $\alpha>0$ the coupling strength. The first term, $-\Delta$ (with $\Delta = \nabla^2$ the Laplacian on $\R^3$), corresponds to the electron kinetic energy, and acts non-trivially only on the electron Hilbert space $L^2(\R^3)$, i.e., it has to be understood as $-\Delta \otimes \id_\F$. The last term is the field energy, which is simply the number operator and corresponds to the energy of a system of uncoupled harmonic oscillators, and likewise has to be understood as $\id_{L^2(\R^3)} \otimes \N$. The  interaction term couples the two systems; it  is an operator on $\F$ indexed by $x$, which is understood as a multiplication operator on the electron Hilbert space $L^2(\R^3)$. 

For any fixed  $x$, the interaction term in \eqref{def:ham} is of the form 
\beq
-a (v_x) - a^\dagger(v_x)
\eeq
where $\hat v_x(k) = \sqrt{\alpha/(2\pi^2)} |k|^{-1} e^{-ik\cdot x}$, i.e., $v_x(y) = \sqrt{\alpha/\pi^3}  |x-y|^{-2}$. This function is not 
 in $L^2(\R^3)$ unless one introduces an ultraviolet cutoff, i.e., restricts the integration to $|k|< \Lambda$ for some $\Lambda>0$. Due to the presence of $-\Delta$, a cutoff $\Lambda$ is not necessary, however. One can define $H_\alpha$ as a quadratic form, and show that the latter is closed and bounded from below (see Subsect.~\ref{ss:LY}), which thus defines naturally $H_\alpha$ as the corresponding operator.  Its operator  domain is complicated, however, and does not contain any finite-phonon vectors, see e.g. \cite{GW}. In particular, $H_\alpha$ is not defined on the domain of $H_0 = -\Delta + \N$. The domain of $H_\alpha$ can be identified via a Gross transformation (explained in Subsect.~\ref{ss:gross} below).

We note that similar models of the kind above, i.e., Hamiltonians of the form 
\beq
H = -\Delta - \int_{\R^3} dk\, v(k) \left( a_k e^{ik\cdot x} + a^\dagger_k e^{- ik\cdot x} \right) + \int_{\R^3} dk\,  \omega(k) a^\dagger_k a_k 
\eeq
for different functions $v$ and $\omega$ 
appear in many places in physics,  and are used as toy models of quantum field theory. 
Examples include, e.g.,
\begin{itemize}
\item the Nelson model of quantum electrodynamics, where $v(k) \propto |k|^{-1/2}$ and $\omega(k) \propto |k|$ or, more generally, $\sqrt{k^2 + m^2}$ for $m\geq 0$ \cite{nelson}.
\item spin-boson models, where $L^2(\R^3)$ for the particle is replaced by $\C^N$, and the coupling  $e^{-ik\cdot x}$ by some function $\R^3\to \M_{N}$, the complex-valued $N\times N$ matrices (see, e.g.,  \cite{spinboson}). 
\item the angulon model, where $L^2(\R^3)$ for the particle is replaced by $L^2(\Sph^2)$, and $e^{-ik\cdot x} $ by the spherical harmonics $Y_{\ell,m}(\Omega)$ \cite{angulon}.
\end{itemize}

\subsection{Strong Coupling and Classical Approximation}

We shall now explain how the case of strong coupling $\alpha \gg 1$ in \eqref{def:ham} leads to a classical approximation of the Fr\"ohlich polaron model. More precisely, only the field variables are treated classically, the electron is still quantum. In this classical approximation, the problem of the ground state (and corresponding ground state energy) can be solved exactly. 

In order to see the emergence of a classical limit, we change variables in the form 
\beq
x \to \alpha^{-1} x \ , \quad a_k \to  \alpha^{-1/2} a_{\alpha^{-1} k} \,.
\eeq
A simple calculation shows that in the new variables, $H_\alpha$ is of the form $\alpha^2 \ch_\alpha$, i.e., $\alpha^{-2} H_\alpha \cong \ch_\alpha$, with
\beq\label{def:ham:sc}
\ch_\alpha := -\Delta -  \frac 1{\sqrt{2\pi^2}} \int_{\R^3}  \frac {dk} {|k|} \left( a_k e^{i k \cdot x} + a_k^\dagger e^{-ik\cdot x} \right)  + \int_{\R^3}dk\,  a_k^\dagger a_k \,.
\eeq
This looks simply like the original Fr\"ohlich Hamiltonian \eqref{def:ham} for $\alpha=1$. Note, however, that the CCR for the transformed creation and annihilation operators 
are now 
\beq
[a_k , a_l^\dagger] = \alpha^{-2} \delta(k-l) \quad , \quad [a_k , a_l^\dagger] = 0 \,.
\eeq
That is, the $\alpha$-dependence is in the CCR. In particular, $\alpha^{-2}$ plays the role of an effective Planck constant, and hence $\alpha\to\infty$ corresponds to a classical limit in which the field operators commute. 

The classical approximation thus amounts to replacing $a_k$ by a complex-valued function $z_k$, and likewise $a^\dagger_k$ by its complex conjugate $z_k^*$. We write $z_k$  as a Fourier transform 
\beq
z_k =  (2\pi)^{-3/2} \int_{\R^3} dx \left( \varphi(x) + i \pi(x) \right) e^{-ik\cdot x} 
\eeq
where $\varphi$ and $\pi$ are real-valued. After this replacement, and taking the expectation in an electron wave functions $\psi\in L^2(\R^3)$, we obtain 
 the Pekar energy functional \cite{P54}
\begin{align}\nonumber
&\E(\psi,\varphi,\pi) \\ \nonumber 
& = \int_{\R^3} dx\, |\nabla\psi(x)|^2  - \frac{\sqrt{2}}{\pi} \Re  \iint_{\R^6} dx\,dk\,  \frac 1{|k|}  |\psi(x)|^2 e^{ik\cdot x} z_k    + \int_{\R^3} dk\, |z(k)|^2 
\\
&= \int_{\R^3} dx\,  |\nabla\psi(x)|^2 - \frac 2 { \pi^{3/2}} \iint_{\R^6} dx\, dy\, \frac{ \varphi(x) |\psi(y)|^2}{|x-y|^2} + \int_{\R^3} dx \left(   |\varphi(x)|^2 + |\pi(x)|^2 \right)  \,. \label{def:pekar3}
\end{align}

Minimizing \eqref{def:pekar3} with respect to $\varphi$ and $\pi$ for fixed $\psi$ leads to the choice 
\beq
\pi(x)=0 \quad, \quad \varphi(x) = \pi^{-3/2} \int_{\R^3} dy \frac{|\psi(y)|^2}{|x-y|^2} 
\eeq
and as energy the corresponding  functional (also called the Pekar functional)
\beq\label{def:Ep}
\E^{\rm Pek}(\psi) = \min_{\varphi,\pi} \, \E(\psi,\varphi,\pi) = \int_{\R^3} dx\,  |\nabla\psi(x)|^2  -  \iint_{\R^6} dx\, dy\,  \frac{ |\psi(x)|^2 |\psi(y)|^2 }{|x-y|} 
\eeq
where we used that
\beq
\frac 1{|x-y|} = \frac 1{\pi^{3}} \int_{\R^3} dz\, \frac 1{|x-z|^2} \frac 1{|y-z|^2} \,.
\eeq
The interaction with the  polarization field thus leads to an effective Coulomb self-interaction of the electron. 

Concerning the minimization with respect to $\psi$, the following Theorem holds. 

\begin{theorem}[Lieb 1977 \cite{Lieb}]\label{thm1}
There exists a minimizer of $\E^{\rm Pek}$ (subject to $\|\psi\|_2 =1$), and it is unique up to translations and multiplication by a phase. That is, every minimizer is of the form
\beq
\psi^{\rm Pek}(x-y) e^{i\theta}
\eeq
for $\theta \in [0,2\pi)$, $y\in \R^3$.
\end{theorem}

In particular, the classical approximation leads to self-trapping of the electron due to its interaction with the polarization field, i.e., the existence of a minimizer despite the translation-invariance of the system. In contrast, in the quantum case, i.e., the full Fr\"ohlich Hamiltonian \eqref{def:ham}, self-trapping is not expected to occur  and ground states do not exist \cite{GL,spohnD}.

We note that the uniqueness part of Theorem~\ref{thm1} is far from obvious, due to the lack of convexity, and the proof in \cite{Lieb} (see also \cite{MZ,TM})  is very specific to the Coulomb interaction. A corresponding result for general interaction potentials is not known, except for small perturbations of either the potential \cite{ricaud} or the nonlinearity \cite{xiang}.

\subsection{Stability}\label{ss:LY}
In this section we show that the Fr\"ohlich Hamiltonian $H_\alpha$ in \eqref{def:ham} is (in the sense of quadratic forms) bounded from below, following the argument by Lieb and Yamazaki in \cite{LY}. 

For $K>0$, write
\beq\label{lycomm}
\int_{|k|>K} \frac {dk}{|k|} a_k e^{ik\cdot x} = \sum_{j=1}^3 \left [ p_j , \int_{|k|>K} dk \frac {k_j}{|k|^3} a_k e^{ik\cdot x} \right]
\eeq
where $p = -i\nabla$ denotes the electron momentum operator. 
With the aid of the Cauchy--Schwarz inequality, we can thus bound
\begin{equation}\label{eq:ly1}
\int_{|k|>K} \frac {dk}{|k|} \left(  a_k e^{ik\cdot x} + a^\dagger_k e^{-ik\cdot x} \right) \leq 2\epsilon (-\Delta) + \frac 1\epsilon \sum_{j=1}^3 \left( a^\dagger(\chi_x^j) a(\chi_x^j) + a(\chi_x^j) a^\dagger(\chi_x^j) \right)
\end{equation}
for any $\epsilon>0$, 
where $\hat\chi_x^j(k) = k_j |k|^{-3} e^{-ik\cdot x} \chi_{|k|>K}$.
Using the CCR and the fact that the functions $\hat\chi_x^j$ are orthogonal and have the same norm, we  can further bound this as
\beq
\eqref{eq:ly1} \leq 2\epsilon (-\Delta) + \frac 2\epsilon \N \|\chi_x^j\|^2 + \frac 1\epsilon \sum_{j=1}^3 \|\chi_x^j\|^2\,.
\eeq
Since
\beq
\|\chi_x^j\|^2 = \frac 13 \sum_{j=1}^3 \|\chi_x^j\|^2 = \frac 13 \int_{|k|>K} \frac {dk}{|k|^4} = \frac {4\pi}3 \frac 1K
\eeq
we can choose $\epsilon = 8\pi  \sqrt{\alpha}(2\pi^2)^{-1/2} (3 K)^{-1}$ and obtain
\begin{align}\nonumber
H_\alpha & \geq \left( 1 - \frac {8{\alpha}}{3\pi K}\right) (-\Delta) - \sqrt{\frac \alpha {2\pi^2}} \int_{|k|<K} \frac {dk}{|k|} \left( a_k e^{ik\cdot x} + a^\dagger_k e^{-ik\cdot x} \right)  \\ & \quad  + \int_{|k|<K} dk\,  a^\dagger_k a_k  - \frac 32 \label{1.24}
\end{align}
for any $K>0$. 
In particular, by choosing $K= 8{\alpha}/(3\pi)$, we obtain
\beq\label{rlb}
H_\alpha\geq - \frac {16}{3\pi^2} \alpha^2 - \frac 32\,.
\eeq
To arrive at this last bound, we have used that 
\beq
a(g) + a^\dagger(g) + \N \geq a(g) + a^\dagger(g) + \|g\|^{-2} a^\dagger(g) a(g) \geq - \|g\|^2
\eeq
for any $g\in \Hh_1$, and applied it to $g(k) = -\sqrt{\alpha/(2\pi^2)} |k|^{-1} e^{-ik\cdot x} \chi_{|k|<K}$. The bound \eqref{rlb} shows, in particular, that $H_\alpha$ is bounded from below, and its ground state energy decreases at most like $-O(\alpha^2)$ for large $\alpha$. 

We remark that the above bounds can be easily generalized to show that $(1+\epsilon) H_0 + C_\epsilon \geq H_\alpha \geq (1-\epsilon) H_0 - C_\epsilon$ for any $\epsilon>0$ and a suitable ($\alpha$-dependent) constant $C_\epsilon>0$. In particular, the form domains of $H_\alpha$ and $H_0$ coincide. 

\subsection{Gross Transformation}\label{ss:gross}

Another way to see that $H_\alpha$ is bounded from below, which in addition identifies also the operator domain of $H_\alpha$, is to apply a Gross transformation \cite{gross,nelson}. 
We write  again
\beq
H_\alpha =-\Delta - a(v_x) - a^\dagger(v_x) + \N
\eeq
and consider a unitary transformation of the form
\beq
U = e^{a(f_x) - a^\dagger(f_x)}
\eeq
for some $f_x\in \Hh_1$ parametrized by $x\in \R^3$. One checks that 
\beq
U a(g) U^\dagger = a(g) + \langle g|f_x\rangle
\eeq
and
\beq
U p U^\dagger = p + a^\dagger(pf_x) + a(pf_x) + \underbrace{ \Re \langle f_x| pf_x\rangle}_{=0 \text{\ for real-valued $f$}}
\eeq
where $p=-i\nabla_x$ is the electron momentum operator, and $pf_x$ stands for $(-i)$ times the gradient of $f_x$ with respect to $x$. This leads to 
\begin{align}\nonumber
UH_\alpha U^\dagger & = p^2 + \left( a^\dagger(pf_x) + a(pf_x)\right)^2 + 2 p\cdot a(pf_x) + 2 a^\dagger(pf_x) \cdot p \\ & \quad + a(p^2 f_x + f_x - v_x) + a^\dagger(p^2 f_x + f_x - v_x) + \N + \|f_x\|^2 - 2 \Re \langle v_x| f_x\rangle\,.
\end{align}
We choose $f_x\in L^2(\R^3)$ such that $pf_x \in L^2(\R^3)$, $|\langle v_x| f_x \rangle| <\infty$ and 
\beq
p^2 f_x + f_x - v_x \in L^2(\R^3) \,.
\eeq
In particular, $p^2 f_x \not\in L^2(\R^3)$ since $v_x  \not\in L^2(\R^3)$. A possible choice for $f_x$ is 
\beq
\hat f_x(k) = \frac 1{k^2+1} \hat v_x(k) \chi_{|k|>K} = \sqrt{\frac \alpha {2\pi^2}} \frac 1{k^2+1} \frac 1 {|k|} e^{-ik\cdot x} \chi_{|k|>K} 
\eeq
for some $K\geq 0$. 
By choosing $K$ appropriately, one can conclude (we refer to~\cite{GW} for details)  that
\beq
(1+ \epsilon) \| H_0 \Psi\| + C_\epsilon \|\Psi\| \geq \| UH_\alpha U^\dagger \Psi \| \geq (1-\epsilon) \| H_0 \Psi\| - C_\epsilon \|\Psi\|
\eeq
for any $\epsilon>0$ and a suitable ($\alpha$-dependent) constant $C_\epsilon$. 
In particular, with $K$ chosen  large enough such that the corresponding $\epsilon <1$, the operator domain of $UH_\alpha U^\dagger$ equals the one of $H_0$.

\subsection{Strong Coupling Limit}\label{ss:SCL}

Recall the definition \eqref{def:Ep} of the Pekar energy functional $\E^{\rm Pek}$, and let $e^{\rm Pek}<0$ denote the Pekar energy
\beq
e^{\rm Pek} = \min_{\|\psi\|_2=1} \E^{\rm Pek}(\psi)\,.
\eeq
Let $e_\alpha$ denote the ground state energy of $H_\alpha$, rescaled  by a factor $\alpha^{-2}$ for simplicity:
\beq
e_\alpha = \alpha^{-2} \infspec H_\alpha =  \infspec \ch_\alpha\,.
\eeq
It is easy to see that $e_\alpha \leq e^{\rm Pek}$ for any $\alpha>0$. To obtain this upper bound, one chooses a trial state of the product form $\psi\otimes \Phi$ for $\psi \in L^2(\R^3)$, $\Phi\in\F$ to get
\beq
\alpha^2 e_\alpha \leq \int_{\R^3} |\nabla\psi|^2 - \int_{\R^3} dx\,  |\psi(x)|^2 \langle \Phi | a(v_x) + a^\dagger(v_x) | \Phi\rangle_\F  + \langle \Phi | \N | \Phi|\rangle_\F \,.
\eeq
For any fixed $\psi$, the optimal choice of $\Phi$ is a coherent state, 
\beq
\Phi = e^{a(g) -a^\dagger(g)} |0\rangle
\eeq
where $g = \int dx\, |\psi(x)|^2 v_x $ and $|0\rangle$ denotes the Fock space vacuum. This choice of $\Phi$ yields the upper bound 
\beq
\alpha^2 e_\alpha \leq \E^{\rm Pek}_\alpha(\psi) = \int_{\R^3} |\nabla \psi|^2 - \alpha \iint_{\R^6} dx\,dy\, \frac{|\psi(x)|^2 |\psi(y)|^2}{|x-y|} \,.
\eeq
By scaling, the infimum over $\psi$ of the right hand side equals $\alpha^2 e^{\rm Pek}$. 

The following quantitative lower bound was proved by Lieb and Thomas in \cite{LT}.

\begin{theorem}[Lieb and Thomas 1997 \cite{LT}]\label{thm:LT}
As $\alpha\to \infty$, 
\beq
e_\alpha =  \infspec \ch_\alpha \geq  e^{\rm Pek} -  O(\alpha^{-1/5}) \,.
\eeq
\end{theorem}

In particular, in combination with the upper bound derived above, we have
\beq
\lim_{\alpha \to \infty} e_\alpha = e^{\rm Pek}\,.
\eeq
 This limit statement  was proved earlier by Donsker and Varadhan in \cite{DV} (see also \cite{leschke}). They used Feynman's \cite{feynman} path integral formulation of the problem, leading to a study of the path measure
\beq
\exp\left( \alpha\int_\R ds \frac{e^{-|s|}}{2} \int_0^T \frac {dt}{| \omega(t) - \omega(t+s)|}\right) d \mathbb{W}^T(\omega)
\eeq
as $T\to \infty$, where $\mathbb{W}^T$ denotes the Wiener measure of closed paths of length $T$. (See also \cite{MV1,MV2} for the construction of the corresponding Pekar process \cite{spohn}.)

Lieb and Thomas  used operator techniques to obtain Theorem~\ref{thm:LT}. We shall sketch their proof in the following.

\begin{proof}[Sketch of proof]
As already shown in \eqref{1.24} above, we have 
\begin{align}\nonumber
H_\alpha & \geq \left( 1 - \frac {8\alpha}{3\pi K}\right) (-\Delta) + \sqrt{\frac\alpha{2\pi^2}} \int_{|k|<K} \frac {dk}{|k|} \left( a_k e^{ik\cdot x} + a_k^\dagger e^{-ik\cdot x} \right)  \\ & \quad + \int_{|k|<K} dk\, a^\dagger_k a_k  - \frac 32 \label{new124}
\end{align}
for any $K>0$. We shall choose $K$ large, $K\sim \alpha^{6/5}$.

We use the IMS localization technique in order to  localize the electron in a cube of side length $L \sim \alpha^{-9/10}$, with a localization error of the order $L^{-2} \sim \alpha^{9/5}$: Let $\Psi\in L^2(\R^3) \otimes \F$ be normalized, and let $E=\langle \Psi | H_\alpha  |\Psi\rangle$. Given $\delta>0$, we claim that there exists a $\phi\in L^2(\R^3)$, supported in some cube of side length $L=\pi \sqrt{2/\delta}$, such that
\beq
\frac{ \langle \phi \Psi | H_\alpha | \phi \Psi\rangle}{\langle \phi \Psi  | \phi \Psi\rangle } \leq E + \delta\,.
\eeq
To see this, simply choose $\phi(x) = \prod_{j=1}^3 \cos(\sqrt{\delta/3} x_j)$ and note that, with $\phi_y(x) = \phi(x-y)$, 
\beq
\int_{\R^3}  dy 
\left( \langle \phi_y \Psi | H_\alpha | \phi_y \Psi\rangle - (E+\delta) \|\phi_y \Psi\|^2 \right) = \int_{\R^3} | \nabla\phi |^2 - \delta \int_{\R^3} |\phi|^2 = 0 \,,
\eeq
hence there must exist a $y\in \R^3$ such that the integrand on the left hand side is non-positive.

Next, we approximate the term $\int_{|k|<K} dk \, |k|^{-1}  a_k e^{ik\cdot x}$ in \eqref{new124}  by a finite sum of the form $\sum_j  \lambda_j a(\varphi_j) e^{i k_j\cdot  x} $ with the $\varphi_j$ orthonormal. In order to do this, we divide momentum space into small cubes $\{C_j\}$ of size $\eps>0$, and take $k_j$ to be the center of the cube $C_j$. Moreover, $\varphi_j$ is proportional to the characteristic function of $C_j$, with $\lambda_j \int \varphi_j = \int_{C_j} dk\, |k|^{-1}$. This introduces only a small  relative error if $L \epsilon$ is small. We shall choose $\epsilon \sim \alpha^{3/5}$, thus $L\epsilon \sim \alpha^{-3/10}$, and a simple estimate borrowing a fraction $\alpha^{-1/5}\N$ from the field energy of the Hamiltonian shows that the error terms can be bounded by $O(\alpha^{9/5})$. 

This approximation reduces the problem to the study of only finitely many phonon modes. In fact, the number $N$ of independent phonon modes is of the order $N\sim (K/\epsilon)^3 \sim \alpha^{9/5}$. The effective Hamiltonian we are led to study is of the form
\beq\label{ledto}
-\Delta + \sum_{j=1}^N \lambda_j \left(    a(\varphi_j) e^{i k_j\cdot  x} + a^\dagger(\varphi_j) e^{-i k_j\cdot x} \right) + \sum_{j=1}^N a^\dagger(\varphi_j) a(\varphi_j) \,.
\eeq
To obtain a lower bound on its ground state energy, one can use the method of upper symbols and coherent states. With $|\vec z\rangle$ for $\vec z \in \C^N$ the eigenstates of $a(\varphi_j)$ corresponding to eigenvalues $z_j \in \C$, we have 
\beq
\id =  \int_{\C^N} dz \, |\vec z\rangle\langle \vec z| \ , \quad  a_j = \int_{\C^N} dz \, z_j |\vec z\rangle\langle \vec z| \ , \quad a^\dagger_j = \int_{\C^N} dz \, z_j^* |\vec z\rangle\langle \vec z|
\eeq
with $dz = \prod_{j=1}^N \frac{ dx_j\, dy_j}{\pi}$ for $z_j = x_j + i y_j$. Moreover, 
\beq
a^\dagger a_j = \int_{\C^N} dz \, \left( | z_j |^2 - 1 \right) |\vec z\rangle\langle \vec z| 
\eeq
and the additional $-1$ in the integrand  leads to an error term for each of the $N$ modes. In total, we have
\beq
\eqref{ledto} = \int_{\C^N} dz \, \left( -\Delta + \sum_{j=1}^N \lambda_j \left(    z_j e^{i k_j\cdot  x} + z_j^* e^{-i k_j\cdot x} \right) + |\vec z|^2  - N \right) |\vec z\rangle\langle \vec z|
\eeq
and for a lower bound we can take the infimum of the spectrum of the operator in parentheses and minimize the latter over $\vec z$. This effectively leads to the classical Pekar problem. 
\end{proof}

Theorem~\ref{thm:LT} shows that the difference between the ground state energy of the Fr\"ohlich Hamiltonian $H_\alpha$ and its classical approximation $\alpha^2 e^{\rm Pek}$ is at most $O(\alpha^{9/5})$. This is presumably far one optimal, it is expected that the true error is in fact $O(1)$ for large $\alpha$, as we shall explain next. This $O(1)$ correction is due to quantum fluctuations of the field about its classical value. 

Let us rewrite the Fr\"ohlich Hamiltonian \eqref{def:ham:sc} in strong coupling units as 
\beq
\ch_\alpha = -\Delta + V_\varphi(x) + \N
\eeq
where
\beq
V_\varphi(x) = - \pi^{-3/2}   \int_{\R^3} dy \frac 1{|x-y|^2} \big( \underbrace{ a_y + a^\dagger_y}_{2\varphi(y)} \big) \,.
\eeq
Since $[\varphi(y),\varphi(y')] =0$ for all $y,y'\in \R^3$, we can choose a representation of Fock space (called \lq\lq Q-space\rq\rq\ in \cite{RS2}) where all the $\varphi(x)$ act as multiplication operators. In particular, we can think of 
\beq
\kappa(\varphi) = \min_{\psi, \pi}  \left( \E(\psi,\varphi,\pi) - \int_{\R^3} \varphi^2 \right)  = \infspec  \left( -\Delta  + V_\varphi(x) \right)  
\eeq
as a multiplication operator on Fock space. 

Note that $\kappa(\varphi)+ \|\varphi\|^2 = \F^{\rm Pek}(\varphi)  \geq e^{\rm Pek}$. Let $H^{\rm Pek} $ denote the Hessian of this functional at a minimizer $\varphi^{\rm Pek}$, i.e., 
\beq\label{Fapp}
\F^{\rm Pek}(\varphi) = e^{\rm Pek} + \langle  \varphi - \varphi^{\rm Pek} | H^{\rm Pek} | \varphi - \varphi^{\rm Pek}\rangle + O (\|\varphi-\varphi^{\rm Pek}\|_2^3) 
\eeq
for $\varphi$ close to $\varphi^{\rm Pek}$. Clearly $H^{\rm Pek}\geq 0$ since we are expanding around the minimum, and also $H^{\rm Pek} \leq \id$ since $\kappa(\varphi)$ is concave in $\varphi$. The Hessian  $H^{\rm Pek}$ has three zero-modes resulting from the translation invariance of the problem, and it is in fact known that these are the only  zero-modes \cite{Lenzmann} (see also \cite{weiwin}). 

The expansion \eqref{Fapp} suggests to approximate
\beq
 \ch_\alpha \approx e^{\rm Pek} + \langle  \varphi - \varphi^{\rm Pek} | H^{\rm Pek} | \varphi - \varphi^{\rm Pek}\rangle - \|\varphi\|^2 + \N \,.
\eeq
The right hand side is simply a system of harmonic oscillators whose ground state energy can be calculated explicitly. Recalling the commutation relation $[a_k,a^\dagger_l]=\alpha^{-2}\delta(k-l)$, one finds $e^{\rm Pek} +  \frac 1 {2\alpha^{2}}  \Tr \left( \sqrt{H^{\rm Pek}} - \id\right)$. One is thus led to the following conjecture.

{\bf 
\begin{conjecture}\label{conj} 
{\normalfont \it As $\alpha\to \infty$, }
\beq\label{eq:conj}
\infspec \ch_\alpha =   e^{\rm Pek} +  \frac 1 {2\alpha^{2}}  \Tr \left( \sqrt{H^{\rm Pek}} - \id\right) + o(\alpha^{-2}) \,.
\eeq
\end{conjecture}
}

The correction term is simply the sum of the ground state energies of many harmonic oscillators (and contains a $-3/(2\alpha^2)$ from the zero modes). Note that it is negative, since $H^{\rm Pek}\leq \id$. It is not difficult to see that $\id - H^{\rm Pek}$ is trace class, hence the trace is well defined and finite. 

The prediction \eqref{eq:conj} is well-known in the physics literature, see \cite{Allcock,Alcock,Gross,tja}. 
It is an open problem to verify it rigorously, however. In the recent work \cite{Frank}, Conjecture~\ref{conj}  was proved for a simplified model where the polaron is confined to a finite region in space and translation invariance is broken. We shall explain this result in the next section.

\section{Quantum Fluctuations for a Confined Polaron}

In this section we shall explain the recent proof in \cite{Frank} of the analogue of Conjecture~\ref{conj} for a confined polaron.  For $\Omega\subset \R^3$ a bounded domain, the Hamiltonian of this model is
\beq\label{haO}
\ch_{\alpha,\Omega}=  -\Delta_\Omega -  \int_{\Omega} dy\,  (-\Delta_\Omega)^{-1/2}(x,y)  \left( a_y + a_y^\dagger \right)  + \int_{\Omega} dy\,  a_y^\dagger a_y 
\eeq
acting on the Hilbert space $L^2(\Omega) \otimes \F(L^2(\Omega))$. Here $\Delta_\Omega$ denotes the Laplacian on $\Omega$ with Dirichlet boundary conditions. We work directly in strong coupling units, i.e., the CCR are $[a_x,a^\dagger_y]=\alpha^{-2} \delta(x-y)$. Note that for $\Omega=\R^3$, the operator \eqref{haO} reduces to $\ch_\alpha$ in \eqref{def:ham:sc} (except for a factor $\sqrt{4\pi}$ in front of the interaction term). 

The classical approximation corresponding to \eqref{haO} is the functional
\begin{align}\nonumber
\E_\Omega(\psi,\varphi,\pi) & = \int_\Omega |\nabla\psi|^2 - 2 \iint_{\Omega\times \Omega} dx\, dy\, |\Psi(x)|^2 (-\Delta)^{-1/2}(x,y) \varphi(y) \\ & \quad + \int_\Omega \left( \varphi^2 + \pi^2\right) 
\end{align}
for $\psi\in H_0^1(\Omega)$ and  $\varphi, \pi \in L^2_\R(\Omega)$. 
The resulting Pekar energy functional  obtained by minimizing over $\varphi$ and $\pi$ is thus
\beq
\E^{\rm Pek}_\Omega(\psi) = \int_\Omega dx\,  |\nabla\psi(x)|^2  - \iint_{\Omega^2} dx\, dy\, |\psi(x)|^2 (-\Delta_\Omega)^{-1}(x,y) |\psi(y)|^2 \,,
\eeq
with corresponding Pekar energy 
\beq
e^{\rm Pek}_\Omega = \min\left\{ \E^{\rm Pek}_\Omega(\psi) \, : \, \psi \in H_0^1(\Omega), \, \|\psi\|_2 = 1\right\}\,.
\eeq

In the following, we need to assume that $\E^{\rm Pek}_\Omega$ has a unique minimizer $\psi^{\rm Pek}_\Omega$ that is non-degenerate, in the sense that 
\beq\label{Pek:as}
\E_\Omega^{\rm Pek}(\psi) \geq \E^{\rm Pek}_\Omega(\psi^{\rm Pek}_\Omega) + K_\Omega \min_\theta  \int_\Omega   | \nabla( \psi - e^{i\theta} \psi^{\rm Pek}_\Omega)|^2
\eeq
for some $K_\Omega>0$, for all $\psi \in H_0^1(\Omega)$ with $\|\psi\|_2=1$. Given that this is known to hold for $\Omega=\R^3$ (in which case one also has to minimize over translations of the Pekar minimizer due to translation invariance), \eqref{Pek:as} is expected to hold for generic domains $\Omega$. It can be proved if $\Omega$ is a ball, see  \cite{FS19}. 

Under the assumption \eqref{Pek:as}  (and some additional regularity assumptions on the boundary of $\Omega$) the following holds.

\begin{theorem}[Two-term Asymptotics of the Ground State Energy \cite{Frank}] \label{thm:FS}
As $\alpha\to \infty$, one has
\beq\label{thm:fs19}
\infspec \ch_{\alpha,\Omega} =   e^{\rm Pek}_\Omega +  \frac 1 {2\alpha^{2}}  \Tr \left( \sqrt{H^{\rm Pek}_\Omega} - \id\right) + o(\alpha^{-2})\,.
\eeq
\end{theorem}

Here $H^{\rm Pek}_\Omega$ denotes the Hessian of $\F^{\rm Pek}_\Omega$ at the unique minimizer $\varphi^{\rm Pek}_\Omega$,  i.e.,
\beq
\langle \varphi | H^{\rm Pek}_\Omega | \varphi\rangle = \lim_{\epsilon\to 0} \epsilon^{-2} \left( \F^{\rm Pek}_\Omega(\varphi^{\rm Pek}_\Omega + \epsilon \varphi) - e^{\rm Pek}_\Omega\right)
\eeq
where
\beq
\F^{\rm Pek}_\Omega(\varphi) = \min_{\psi,\pi} \E_\Omega(\psi,\varphi,\pi) = \infspec \left( -\Delta_\Omega + V_\varphi(x) \right) + \int_\Omega \varphi^2
\eeq
with $V_\varphi(x) = -2 \int_\Omega dy\, (-\Delta_\Omega)^{-1/2}(x,y) \varphi(y)$. 

With the aid of second order perturbation theory, one can obtain an explicit formula for $H^{\rm Pek}_\Omega$. With  $\varphi^{\rm Pek}_\Omega$ the unique minimizer of $\F^{\rm Pek}_\Omega$, one has
\beq\label{2nd}
\left. \frac 12 \frac{d^2}{d\epsilon^2} \infspec \left( -\Delta + V_{\varphi^{\rm Pek}_\Omega+\epsilon \varphi} \right) \right|_{\epsilon = 0}  = - \langle \psi^{\rm Pek}_\Omega | V_{\varphi} \frac Q{-\Delta_\Omega + V_{\varphi^{\rm Pek}_\Omega} - \mu^{\rm Pek}} V_\varphi | \psi^{\rm Pek}_\Omega\rangle 
\eeq
where $\mu^{\rm Pek} = e^{\rm Pek}_\Omega - \|\varphi^{\rm Pek}_\Omega\|_2^2$ denotes the ground state energy of $-\Delta_\Omega + V_{\varphi^{\rm Pek}_\Omega}$, and $Q = 1-|\psi^{\rm Pek}_\Omega\rangle \langle \psi^{\rm Pek}_\Omega|$ is the projection orthogonal to the corresponding ground state. Reordering the terms and using that $V_{\varphi} = -2 (-\Delta_\Omega)^{-1/2}\varphi$ one can alternatively write this as
\beq
\eqref{2nd} = - 4 \langle \varphi | (-\Delta_\Omega)^{-1/2} \psi^{\rm Pek}_\Omega \frac Q{-\Delta_\Omega + V_{\varphi^{\rm Pek}_\Omega} - \mu^{\rm Pek}} \psi^{\rm Pek}_\Omega (-\Delta_\Omega)^{-1/2} | \varphi\rangle
\eeq
where $\psi^{\rm Pek}_\Omega$ is understood as a multiplication operator. In particular, 
 $H^{\rm Pek}_\Omega = \id - K$ with 
\beq
K =  4 (-\Delta_\Omega)^{-1/2} \psi^{\rm Pek}_\Omega \frac Q{-\Delta_\Omega + V_{\varphi^{\rm Pek}_\Omega} - \mu^{\rm Pek}} \psi^{\rm Pek}_\Omega (-\Delta_\Omega)^{-1/2} \,.
\eeq

Before giving the (sketch of the) proof of Theorem~\ref{thm:FS}, we summarize the main ideas as follows.

\begin{itemize}
\item The electron can be considered to always be in the instantaneous ground state of a Schr\"odinger operator with potential  generated by the (fluctuating) phonon field. This  leads to a reduction of the problem to Fock space only. 
\item Because the number of phonon modes is infinite, $\varphi$ cannot be considered to be a function in $L^2(\Omega)$, and it is, in particular, not close to $\varphi^{\rm Pek}$ in $L^2$; in order to localize the field around its classical value, it is necessary to introduce an ultraviolet cutoff $\Lambda$. 
\item The effect of an ultraviolet cutoff in the interaction can be quantified by using the commutator method of Lieb and Yamazaki \cite{LY}, as explained in the previous section. 
\item We apply, in fact, three commutators, and together with a Gross transformation we conclude that the cutoff affects the ground state energy  at most $O(\Lambda^{-5/2})$. 
\item Due to the space confinement present on our model, an ultraviolet cutoff $\Lambda$ effectively makes the number of phonon modes finite, in fact of the order $|\Omega| \Lambda^3$ according to Weyl's law.
\item The final step uses IMS localization in Fock space, resulting in the  Hessian $H^{\rm Pek}_\Omega$ if $\varphi$ is close to $\varphi^{\rm Pek}$. In the opposite regime, a suitable global coercivity bound is needed. 
\end{itemize}

We remark that the spatial confinement is essential for reducing the problem to finitely many phonon modes after the introduction of an ultraviolet cutoff. A similar effect could be obtained by adding a confining potential instead of the spatial restriction to $\Omega\subset \R^3$, and we expect that our method of proof can be adapted  to also deal with this situation if the confining potential increases fast enough at infinity.

\begin{proof}[Sketch of proof]
{\it Step 1.} This first step is to estimate the classical Pekar functional close to its minimum. The following Lemma can be proved with the aid of perturbation theory.

\begin{lemma} If $\| (-\Delta_\Omega)^{-1/2} (\varphi - \varphi^{\rm Pek}_\Omega)\|_2 \leq \epsilon$ for $\epsilon$ small enough, than
\begin{multline}\label{rsop}
\langle \varphi-\varphi^{\rm Pek}_\Omega| \id - (1-\epsilon)K | \varphi-\varphi^{\rm Pek}_\Omega \rangle \\ \geq \F^{\rm Pek}_\Omega(\varphi) - \F^{\rm Pek}_\Omega(\varphi^{\rm Pek}_\Omega) \geq \langle \varphi-\varphi^{\rm Pek}_\Omega|  \id - (1+\epsilon)K | \varphi-\varphi^{\rm Pek}_\Omega \rangle \,.
\end{multline}
\end{lemma}

Our assumption \eqref{Pek:as} can be shown to imply the absence of zero-modes of $H^{\rm Pek}_\Omega$, i.e.,   $\|K\|<1$, hence the right hand side of \eqref{rsop} is positive for $\epsilon$ small enough.

We also need the following rougher global bound, which can be shown to follow from the coercivity assumption \eqref{Pek:as} on $\E^{\rm Pek}_\Omega$. 

\begin{lemma}
For some $\kappa > 0$
\beq\label{lb2}
\F^{\rm Pek}_\Omega(\varphi) - \F^{\rm Pek}_\Omega(\varphi^{\rm Pek}_\Omega )\geq \langle \varphi-\varphi^{\rm Pek}_\Omega| \id - \frac 1{\id+\kappa (-\Delta_\Omega)^{1/2}} | \varphi-\varphi^{\rm Pek}_\Omega \rangle
\eeq
for all $\varphi \in L^2_\R(\Omega)$.
\end{lemma}

{\it Step 2.} For the computation of the upper bound in \eqref{thm:fs19}, we pick a finite-dimensional projection $\Pi$ with range containing $\varphi^{\rm Pek}_\Omega$, and choose a trial state that is non-trivial only in $L^2(\Omega)\otimes \F(\Pi L^2(\Omega))$, and equals the vacuum in $\F((1-\Pi)L^2(\Omega))$. We use again the $Q$-space formulation mentioned in the previous section, where the Fock space is viewed as consisting of functions of $\varphi$.   For $\varphi \in \Pi L^2(\Omega)$, the trial state has the form
\begin{align}\nonumber
\Psi(x,\varphi) & = \exp\left( - \alpha^2 \langle \varphi - \varphi^{\rm Pek}_\Omega | (\id- \Pi K \Pi )^{1/2} | \varphi - \varphi^{\rm Pek}_\Omega \rangle\right) \\ & \quad \times \chi\left( \epsilon^{-1} \| (-\Delta_\Omega)^{-1/2}( \varphi-\varphi^{\rm Pek}_\Omega)\|_2\right) \psi_{\varphi}(x) \label{nts}
\end{align}
where $\epsilon>0$ is a parameter that goes to zero as $\alpha\to \infty$, $\chi$ is a smooth and compactly supported cut-off function that equals one close to the origin, and $\psi_\varphi$ is the ground state of $-\Delta_\Omega + V_\varphi(x)$. The projection $\Pi$ will be chosen in order to converge strongly to $\id$ as $\alpha\to 0$. After optimizing over the choice of $\epsilon$ and $\Pi$, one finds the upper bound
\beq
\frac {\langle \Psi | \ch_{\alpha,\Omega} | \Psi\rangle}{\langle \Psi | \Psi\rangle} \leq e^{\rm Pek} +  \frac 1 {2\alpha^{2}}  \Tr \left( \sqrt{H^{\rm Pek}_\Omega} - \id\right) + O(\alpha^{-24/11})\,.
\eeq 

We remark that the trial state \eqref{nts} differs from the Pekar product ansatz in Section~\ref{ss:SCL} in two essential ways. The first is the appearance of the correction $K$ to the Hessian of $ \F^{\rm Pek}_\Omega$ in the exponent (which is simply taken to be zero  in the Pekar ansatz), and the second is the   function $\psi_\varphi$ in place of its average value $\psi_{\varphi_\Omega^{\rm Pek}} = \psi_\Omega^{\rm Pek}$, which adjusts the electron wave function to the fluctuating phonon field.

{\it Step 3.} A crucial step in the lower bound is to quantify the effect of an ultraviolet (UV) cutoff in the interaction term in the Fr\"ohlich Hamiltonian. Recall the commutator method by Lieb and Yamazaki in \eqref{lycomm}. On the whole space $\R^3$, it reads
\beq
a(v_x) = [ p , a( \nabla(-\Delta)^{-1} v_x )]
\eeq
and is applied with $v_x(k) = |k|^{-1} \chi_{|k|>\Lambda} e^{-i k \cdot x}$. A simple Cauchy--Schwarz inequality shows that the commutator can be bounded by $\sqrt{-\Delta} \sqrt{\N} \|  \nabla(-\Delta)^{-1} v_x\|$. The latter norm is of the order $\Lambda^{-1/2}$, and since both $-\Delta$ and $\N$ are order one (in the strong coupling units we work with here) in low energy states, this shows that the effect of an UV cutoff $\Lambda$ on the ground state energy is at most $O(\Lambda^{-1/2})$. 

We generalize this idea by applying the commutator method three times, arriving at a triple Lieb--Yamazaki bound. Ignoring vector indices for simplicity, we effectively have
\beq
a(v_x) \sim [ p, [p, [p, a( (-\Delta)^{-3/2} v_x] ] ]
\eeq
which can be bounded by $(-\Delta)^{3/2} \sqrt{\N} \| (-\Delta)^{-3/2} v_x\|$, with the latter norm now of the order $\Lambda^{-5/2}$. The operator $(-\Delta)^{3/2} \sqrt{\N}$ can be bounded in terms of $\ch_0^2$, the square of the non-interacting Hamiltonian. The latter is {\em not} bounded in terms of $\ch_\alpha^2$, however, and is in fact infinite in any state in the domain of $\ch_\alpha$. Hence we cannot apply such a bound directly.

What saves the day is the Gross transformation explained in Section~\ref{ss:gross} above. It is a unitary $U$ with the property that $(U^\dagger \ch_0 U)^2 \lesssim \ch_\alpha^2$. Moreover, $U^\dagger a(v_x) U = a(v_x) + \const$ with a constant of the order $\alpha^{-2} \Lambda^{-1}$, which is negligible for our purpose. The conclusion from all this is that an UV cutoff $\Lambda$ can be introduced in the interaction term, at an energy cost of at most $O(\Lambda^{-5/2})$.\footnote{For\label{footn} simplicity we ignore the Dirichlet boundary conditions here. They lead to some technical difficulties which make it easier to work with smooth instead of sharp cutoffs, and lead to an additional logarithmic factor in the analysis in \cite{Frank}.} Hence, as long as $\Lambda \gg \alpha^{4/5}$, we can work with 
\beq\label{haL}
\ch_{\alpha,\Omega}^\Lambda = -\Delta_\Omega - a(v_x^\Lambda) - a^\dagger(v_x^\Lambda) + \N
\eeq
with
\beq
v_x^\Lambda(y) = \frac{ \theta(\Lambda + \Delta_\Omega)}{(-\Delta_\Omega)^{1/2}}(x,y) = \sum_{j=1}^N \sqrt{e_j} \varphi_j(x) \varphi_j(y)
\eeq
where $e_j$ and $\varphi_j$ are the eigenvalues and eigenfunctions of $-\Delta_\Omega$, and $N\sim |\Omega| \Lambda^3$ according to Weyl's law \cite{RS4}. 

{\it Step 4.} We now proceed with a lower bound on the ground state energy of \eqref{haL}. Simply minimizing over the electronic part of the Hamiltonian gives 
\beq\label{rhsS}
\ch_{\alpha,\Omega}^\Lambda \geq \F^{\rm Pek}_\Omega(\varphi) - \frac 1{4\alpha^2} \sum_{j=1}^N \frac{\partial^2}{\partial \lambda_j^2} - \frac N{2\alpha^2}
\eeq
where  $\vec\lambda= (\lambda_1,\dots,\lambda_N) \in \R^N$ are the coefficients in the expansion $\varphi = \sum_{j=1}^N \lambda_j \varphi_j$. 

The right hand side of \eqref{rhsS} is a Schr\"odinger-type operator on $L^2(\R^N)$. 
To obtain a lower bound on its ground state energy, we use IMS localization into a region $\mathfrak{A}$ where $\| (-\Delta_\Omega)^{-1/2}(\varphi-\varphi^{\rm Pek}_\Omega)\|_2 < \epsilon$ and a region $\mathfrak{B}$ where $\|(-\Delta_\Omega)^{-1/2}(\varphi-\varphi^{\rm Pek}_\Omega)\|_2 > \epsilon/2$ for some $\epsilon>0$. The resulting localization error is of the order $\alpha^{-4} \epsilon^{-2}$, hence we need $\epsilon \gg \alpha^{-1}$ for it to be negligible. 

In region $\mathfrak{A}$, we use the second inequality in \eqref{rsop}, which gives the lower bound
\beq 
 e^{\rm Pek}_\Omega - \frac 1 {2\alpha^2} \Tr \left( \id - \sqrt{\id - \Pi K \Pi (1+\epsilon)} \right) \,,
\eeq
which is of the desired form if $\epsilon\to 0$ and $N\to \infty$ as $\alpha\to \infty$. 

In region $\mathfrak{B}$, we use the bound \eqref{lb2} instead, which implies 
\beq\label{lb22}
\F^{\rm Pek}_\Omega(\varphi) \geq e^{\rm Pek}_\Omega + \frac{\eta}4 \epsilon^2 +  \langle \varphi-\varphi^{\rm Pek}_\Omega| \id - \frac 1{\id+\kappa (-\Delta_\Omega)^{1/2}} - \frac \eta {(-\Delta_\Omega)} | \varphi-\varphi^{\rm Pek}_\Omega \rangle
\eeq
in this region for any $\eta >0$. We choose $\eta$ small enough (but independently of $\alpha$) such that the operator whose expectation value is taken on the right hand side of \eqref{lb22} is positive. This gives the lower bound
\beq\label{sec}
e^{\rm Pek}_\Omega + \frac \eta 4 \epsilon^2 - \frac 1{2\alpha^2} \Tr \Pi \left( \id - \sqrt{\id - \frac 1{1+ \kappa (-\Delta_\Omega)^{1/2}} - \frac \eta{(-\Delta_\Omega)} } \right)\,.
\eeq
The latter trace is of the order $N^{2/3} \sim \Lambda^2$. Hence, as long as we choose $\Lambda^2 \alpha^{-2}\leq \const \eps^2$ for a small enough constant, \eqref{sec} is larger than $e^{\rm Pek}_\Omega$. In particular, we need $\Lambda \ll \alpha$, which is compatible with the requirement $\Lambda\gg \alpha^{4/5}$ from above.

After optimizing over the choice of $\Lambda$ and $\epsilon$, we reach the conclusion
\beq
\infspec \ch_{\alpha,\Omega} \geq e^{\rm Pek}_\Omega +  \frac 1 {2\alpha^{2}}  \Tr \left( \sqrt{H^{\rm Pek}_\Omega} - \id\right) - O(\alpha^{-15/7})
\eeq
(modulo logarithmic factors mentioned in the footnote on page \pageref{footn}). 
This completes the (sketch of the) proof of Theorem~\ref{thm:FS}. 
\end{proof}

\section{Effective Mass}

We now return to the full-space Fr\"ohlich Hamiltonian for an unconfined electron on $\R^3$, given (in the original, unscaled variables) by 
\begin{equation}\label{def:hama}
H_\alpha = -\Delta - \sqrt{\frac{\alpha}{2\pi^2}}  \int_{\R^3}  \frac {dk} {|k|} \left( a_k e^{i k \cdot x} + a_k^\dagger e^{-ik\cdot x} \right)  + \int_{\R^3} dk\, a_k^\dagger a_k \,.
\end{equation}
This  Hamiltonian  is translation invariant, i.e., it  commutes with the (three components of the) total momentum operator
\beq
P = -i\nabla_x + P_{\rm f} \ , \quad P_{\rm f} = \int_{\R^3} dk\, k\, a^\dagger_k a_k \,,
\eeq
which follows easily from
\beq
[-i\nabla_x, e^{ik\cdot x}] = k\, e^{ik\cdot x} \ , \quad [ a^\dagger_k a_k , a_l ] = -\delta(k-l) k \, a_k \, . 
\eeq
Hence there is a fiber-integral decomposition 
\beq
H_\alpha = \int_{\R^3}^\oplus dP\, H_\alpha^P \, .
\eeq
In fact, the fiber operators $H_\alpha^P$ are isomorphic to  \cite{LLP}
\beq
H_\alpha^P \cong  \left( P - P_{\rm f} \right)^2 - \sqrt{\frac\alpha {2\pi^2}} \int_{\R^3}  \frac {dk} {|k|} \left( a_k  + a_k^\dagger  \right)  + \int_{\R^3} dk\, a_k^\dagger a_k 
\eeq
acting on $\F$ only.  In following, we shall investigate the joint energy-momentum spectrum.

Let $E_\alpha(P) = \infspec H_\alpha^P$, and let us first consider the non-interacting case $\alpha=0$. The operator $H_0^P$ has a single eigenvalue $P^2$ with the vacuum the corresponding eigenfunction, and a continuous spectrum in $[1,\infty)$. In particular
\beq
E_0(P) = \begin{cases} P^2 & \text {for $|P|\leq 1$} \\ 1 & \text{for $|P|\geq 1$}
\end{cases}
\eeq
and $E_0(P)$ is an eigenvalue if and only if $|P|\leq 1$. 

This picture turns out to be   qualitatively the same for $\alpha>0$ \cite{JSM} (see also \cite{spohnD}). We have $\infspec H_\alpha = E_\alpha(0)$, i.e., $E_\alpha(P)$ is minimal at $P=0$. The continuous spectrum of $H_\alpha^P$ starts at $E_\alpha(0) + 1$, and $E_\alpha(P)$ is  a simple eigenvalue if $E_\alpha(P) < E_\alpha(0) + 1$, which is true for $|P|$ small enough.

The {\em effective mass} $m$ of the polaron is defined by 
\beq
E_\alpha(P) = E_\alpha(0) + \frac{P^2}{2m} + o(|P|^2) \ \text{as $P\to 0$},
\eeq
that is, 
\beq
\frac 1m := 2 \lim_{P\to 0}  \frac{E_\alpha(P) - E_\alpha(0)}{|P|^2}\,.
\eeq
Clearly $E_\alpha(P) \leq E_\alpha(0) + P^2$, hence $m\geq 1/2$ (and the inequality is actually strict for $\alpha>0$). 

Alternatively, one can define the effective mass $m$ via the ground state energy of $H_\alpha + V(x)$ for a slowly varying external potential $V$ of the form $V(x) = \lambda^2 W(\lambda x)$, by comparing with the ground state energy of the corresponding Schr\"odinger operator $-\Delta/m + V(x)$ as $\lambda \to 0$ \cite{LSm}. 

A simple argument based on the Pekar approximation \cite{pekar} suggests that $m \sim \alpha^4$ as $\alpha \to \infty$. To see this, one envisions a slowly moving polaron of the form $\psi(x,t) = \psi^{\rm Pek}(x - v t)$, $\varphi(x,t) = \varphi^{\rm Pek}(x- v t)$, where $\psi^{\rm Pek}$ minimizes the Pekar functional $\E^{\rm Pek}$ in \eqref{def:Ep}  and $\varphi^{\rm Pek}$ denotes the corresponding Pekar field $\varphi^{\rm Pek} = \pi^{-3/2} |\psi^{\rm Pek}|^2 * |x|^{-2}$ (and we work in strong coupling units for simplicity). 
One of the classical equations of motion is $\partial_t \varphi = \alpha^{-2} \pi$, hence the motion requires the additional field energy $\int \pi^2 = \alpha^4\int (\partial_t \varphi)^2 = \alpha^4\int ( v\cdot \nabla \varphi^{\rm Pek})^2$. Identifying this energy with $(m/2) v^2$, one arrives at the 

{\bf 
\begin{conjecture}\label{conj2}
\beq
\lim_{\alpha \to \infty} \alpha^{-4} m = \frac 2 3 \int_{\R^3} |\nabla \varphi^{\rm Pek}|^2 = \frac{8\pi}3 \int_{\R^3} |\psi^{\rm Pek}|^4  \,.
\eeq
\end{conjecture}
}

The best rigorous result so far is
\begin{theorem}[Divergence of the Effective Mass \cite{LiebS19}]\label{thm:ls19}
\beq
\lim_{\alpha\to \infty} m = \infty
\eeq
\end{theorem}

For its proof, we need an upper bound on the difference $E_\alpha(P) - E_\alpha(0)$. This will be achieved with a suitable trial state for $H^P_\alpha$, which is constructed with the aid of the ground state $\Phi_0$ of $H_\alpha^{P=0}$. 

The motivation for the specific choice of the trial state is as follows. As $\alpha \to \infty$, the ground state energy of the Fr\"ohlich Hamiltonian is to leading order captured by the Pekar product ansatz
$\psi \otimes \Phi$, 
where $\Phi$ is a coherent state in $\F$, i.e., is proportional to  $e^{a^\dagger(\varphi)} |0\rangle$ for some $\varphi \in L^2(\R^3)$ and $|0\rangle$ the vacuum state. Decomposing the Pekar  ansatz state into fibers suggests for the fiber ground states $\Phi_P$
\beq
\Phi_P \approx \hat\psi_\alpha^{\rm Pek}(P-P_f) e^{a^\dagger(\varphi_\alpha ^{\rm Pek}) } |0\rangle \approx \Phi_0 + P \cdot \frac{ \nabla \hat\psi_\alpha^{\rm Pek}(-P_f)}{\hat\psi_\alpha^{\rm Pek}(-P_f)} \Phi_0 
\eeq
where the subscript $\alpha$ on $\psi^{\rm Pek}$ and $\varphi^{\rm Pek}$ indicates rescaling to the original variables, i.e., $\hat\psi^{\rm Pek}_\alpha(p) = \alpha^{-3/2} \hat\psi^{\rm Pek}(\alpha^{-1}p)$ and $\varphi^{\rm Pek}_\alpha(p) = \alpha^{-1/2} \varphi^{\rm Pek}(\alpha^{-1}p)$. 
The idea is now to use this as a trial state for $H_\alpha^P$, with $\Phi_0$ the actual ground state of $H_\alpha^0$.

Let\footnote{Since we don't know whether $t$ is bounded (at infinity), we take a mollified version $t_\epsilon$ and let $\epsilon\to 0$ at the end.}
\beq\label{def:t}
t(p) = \frac{\nabla \hat \psi^{\rm Pek}(-p)}{\hat\psi^{\rm Pek}(-p)} 
\eeq
and define
\beq
\Phi_P = \Phi_0 - \alpha^{-1} P \cdot t(\alpha^{-1} P_{\rm f}) \Phi_0
\eeq
with $\Phi_0$ the ground state of $H_\alpha^{P=0}$. By rotation invariance, we have 
\beq
\| \Phi_P\|^2 = 1 + \frac{P^2}{3\alpha} \langle \Phi_0 | t(\alpha^{-1} P_{\rm f})^2 | \Phi_0\rangle\,.
\eeq
Moreover, since $H_\alpha^0 \Phi_0 = E_\alpha(0) \Phi_0$, we further have
\begin{align}\nonumber
\langle \Phi_P | H_\alpha^P | \Phi_P\rangle & = E_\alpha(0) + P^2 + \alpha^{-2} \langle P\cdot t \Phi_0 | H_\alpha^0 | P\cdot t \Phi_0 \rangle \\ & \quad  + 4 \alpha^{-1} \langle \Phi_0 | P\cdot P_{\rm f} P\cdot t | \Phi_0\rangle + O(|P|^3)
\end{align}
where $t$ is short for $t(\alpha^{-1} P_{\rm f})$. 
In particular, 
\begin{align}\nonumber
\frac 1{2m} & \leq  \lim_{P\to 0} \frac 1{|P|^2} \left( \frac { \langle \Phi_P | H_\alpha^P | \Phi_P\rangle}{\| \Phi_P\|^2} - E_\alpha(0) \right) \\ 
& = 
1 + \frac 1{3\alpha^2} \langle t \Phi_0 | H_\alpha^0 - E_\alpha(0)  | t \Phi_0\rangle
+ \frac 4{3 \alpha}  \langle \Phi_0 | P_{\rm f} \cdot t  | \Phi_0\rangle \,. \label{claimr}
\end{align}
 
We claim that the right hand side of \eqref{claimr} goes to zero as $\alpha\to \infty$ if $t$ is chosen as in \eqref{def:t}. 
This claim is an immediate consequence of the following Lemma.

\begin{lemma}\label{lem:LT}
For any function $g$ with bounded second derivative, 
\beq
\lim_{\alpha\to \infty} \langle \Phi_0 | g(\alpha^{-1}P_{\rm f} )| \Phi_0 \rangle = \int_{\R^3} |\hat \psi^{\rm Pek}|^2 g \,.
\eeq
If in addition $g$ is bounded, 
\beq
\lim_{\alpha\to \infty} \alpha^{-2} \langle \Phi_0 | \N g(\alpha^{-1}P_{\rm f}) | \Phi_0 \rangle = \int_{\R^3} |\hat \psi^{\rm Pek}|^2 g \, \int_{\R^3} (\varphi^{\rm Pek})^2
\eeq
and 
\begin{align}\nonumber
& \lim_{\alpha\to \infty} \alpha^{-1} \langle \Phi_0 | g(\alpha^{-1}P_{\rm f} ) a^\dagger(\xi_\alpha) g(\alpha^{-1}P_{\rm f})  | \Phi_0 \rangle  \\ & = \iint_{\R^6} dk\, dp\, \varphi^{\rm Pek}(k) \xi(k) \hat \psi^{\rm Pek*}(p+k) g(p+k) \hat \psi^{\rm Pek}(p) g(p) 
\end{align}
where $\xi_\alpha(p) = \alpha^{-3/2} \xi (\alpha^{-1} p)$. 
\end{lemma}

To prove this Lemma, one follows the Lieb--Thomas proof in \cite{LT} for a perturbed Hamiltonian of the form
\beq
H^0_\alpha(\vec \lambda) = H_\alpha^0 + \lambda_1 \alpha^2 g(\alpha^{-1} P_{\rm f}) + \lambda_2 \dots
\eeq
and shows that
\beq\label{suffl}
\lim_{\alpha\to \infty} \alpha^{-2} \infspec H_\alpha^0(\vec \lambda) = \inf_{\psi,\varphi} \left\{ \E(\psi,\varphi) + \lambda_1 \int_{\R^3} |\hat \psi|^2 g + \lambda_2 \dots\right\}
\eeq
for $|\vec \lambda|$ small enough.  Differentiating this identify with respect to $\vec\lambda$ at $\vec \lambda =0$ gives the desired identities. 
It  is in fact sufficient to prove a  lower bound in \eqref{suffl}, and for that purpose one can re-introduce the electron coordinate and proceed similarly as in \cite{LT}.

From Lemma~\ref{lem:LT} we conclude that 
\begin{align}\nonumber
\lim_{\alpha\to \infty} \frac 4{3\alpha}  \langle \Phi_0 | P_{\rm f} \cdot t(\alpha^{-1}P_{\rm f}) | \Phi_0\rangle & = \frac 4 3 \int_{\R^3} dp\,  | \hat \psi^{\rm Pek}(p)|^2 p\cdot t(p)  \\ &= \frac 43 \int_{\R^3} dp\,  \hat \psi^{\rm Pek}(p) p\cdot \nabla \hat \psi^{\rm Pek}(p)  = -2
\end{align}
and hence Theorem~\ref{thm:ls19} is proven if 
\beq\label{isp}
\lim_{\alpha\to \infty} \alpha^{-2}  \langle t \Phi_0 | H_\alpha^0 - E_\alpha(0)  | t \Phi_0\rangle = 3\,.
\eeq
Applying again Lemma~\ref{lem:LT} and using the fact that $\lim_{\alpha\to\infty}\alpha^{-2} E_\alpha(0) = e^{\rm Pek}$ one can show that  the left hand side of \eqref{isp} equals
\begin{align}\nonumber
&\int_{\R^3} dp \, |\nabla \hat \psi^{\rm Pek}(p)|^2 \left( p^2 + \int_{\R^3} (\varphi^{\rm Pek})^2 - e^{\rm Pek}\right) \\ & - 2 \iint_{\R^6} dp\, dk\, \frac{\varphi^{\rm Pek}(k)}{|k|} \nabla \hat \psi^{\rm Pek}(p+k) \cdot \nabla \hat \psi^{\rm Pek}(p) \,. \label{lse1}
\end{align}
The Euler--Lagrange equation satisfied by $\hat\psi^{\rm Pek}$ reads
\beq\label{gtbt}
\left( p^2 + \mu \right) \hat\psi^{\rm Pek}(p) - 2 \int_{\R^3} dk\, \frac{\varphi^{\rm Pek}(k)}{|k|} \hat\psi^{\rm Pek}(p+k)  = 0
\eeq
for $\mu = \int (\varphi^{\rm Pek})^2 - e^{\rm Pek}$. Taking the gradient of \eqref{gtbt} with respect to $p$, multiplying by $\nabla \hat\psi^{\rm Pek}(p)$ and integrating over $p$ shows that
\beq
\eqref{lse1} = -2 \int_{\R^3} dp\, \hat\psi^{\rm Pek}(p) p\cdot \nabla \hat\psi^{\rm Pek}(p) = 3
\eeq
as claimed.

\section{Conclusion and Open Problems}

In the previous two sections, we explained the quantum corrections to the (classical) Pekar asymptotics of the ground state energy of a confined polaron, and showed that the polaron's effective mass diverges in the strong coupling limit. Many open problems remain, in particular Conjecture~\ref{conj} concerning the second term in the ground state energy of an unconfined polaron in the strong coupling limit, and Conjecture~\ref{conj2} concerning the asymptotics of the effective mass. 

We note that the Pekar approximation can also be applied in the dynamical setting. Viewing \eqref{def:pekar3} as a Hamiltonian system, the corresponding Landau--Pekar equations are expected to approximate the dynamics generated by the Fr\"ohlich Hamiltonian \eqref{def:ham:sc} for suitable initial states. We refer to \cite{LMRSS} for a recent proof of this claim, and also to  
 \cite{frankgang,frankgang3,frankschlein,griesemer,LRSS} for partial results on this topic. 
  
 Finally, we want to mention that the study of polaronic interactions for many particle systems leads to interesting problems concerning the existence of bound states due to the effective attraction via the polarization field, and 
the resulting question of {stability} of the system for large particle number. We refer to 
 \cite{FLST} for an overview of recent work in this direction.

\section*{Acknowledgments}

This work was supported by the European Research Council (ERC) under the European Union's Horizon 2020 research and innovation programme (grant agreement No 694227).


\end{document}